\documentclass{article}

\usepackage{arxiv}
\usepackage{natbib}

\usepackage[utf8]{inputenc} % allow utf-8 input
\usepackage[T1]{fontenc}    % use 8-bit T1 fonts
\usepackage{hyperref}       % hyperlinks
\usepackage{url}            % simple URL typesetting
\usepackage{booktabs}       % professional-quality tables
\usepackage{amsfonts}       % blackboard math symbols
\usepackage{nicefrac}       % compact symbols for 1/2, etc.
\usepackage{microtype}      % microtypography
\usepackage{lipsum}		% Can be removed after putting your text content
\usepackage{amsmath}
\usepackage{amssymb}
\usepackage{graphicx}
\usepackage{caption}
\usepackage{subcaption}
\usepackage[font=small,skip=0pt]{caption}

\title{Inlet swirl decay and mixing in a laminar micropipe flow with wall slip}

\author{
  Dhananjay Kumar \\
    Department of Mechanical Engineering \\
  National Institute of Technology Tiruchirappalli\\
  Tamil Nadu, India – 620015 \\
   \And
 Shavitur Mukesh Kumar Shakhya \\
  Department of Mechanical Engineering \\
  National Institute of Technology Tiruchirappalli\\
  Tamil Nadu, India – 620015 \\
  \And
 P Kaushik\thanks{corresponding author alternate email: kaushik.engg@gmail.com} \\
   Department of Mechanical Engineering \\
  National Institute of Technology Tiruchirappalli\\
  Tamil Nadu, India – 620015 \\
  \texttt{pkaushik@nitt.edu} \\
}

\begin{document}
\maketitle

\begin{abstract}
In this work, the laminar decaying inlet swirl flow in a straight micro-pipe with wall slip is solved analytically and the solution verified numerically. Based on a fully developed parabolic axial velocity profiles, the swirl velocity equation is solved by the separation of variable technique. The solution is expressed as a function of the flow Reynolds number, the axial distance within the micro-pipe from the inlet, the wall slip and the inlet swirl velocity profile. The effects of the parameters on the swirl velocity distribution and the swirl decay are analyzed along the flow. Addition of a swirling velocity to the flow of a fluid in a pipe is of great importance in enhancement transport characteristics. The current results offer analytical equations to estimate the swirl velocity distribution with slip at the walls for the design of swirl flow devices. Furthermore, to quantify mixing, the change in the average distance traveled by fluid particles from inlet in a swirl flow is compared with the average distance traveled by the fluid particles in case of no swirl. A clear enhancement of the average distance traveled is obtained. In our opinion the present work is useful to researchers looking for enhancement of transport characteristics in micro-pipes.
\end{abstract}

% keywords can be removed
\keywords{wall slip \and swirl decay \and inlet swirl \and Laminar pipe flow \and axisymmetric \and swirl number \and transition radius \and Rankine vortex}

\section{Introduction}
As microchannel based fluid flow devices become popular, there is need to understand the transport properties of fluid flow with in such small channels and tubes. One of the traditional methods to enhance transport in tubes is by introduction of swirl [\cite{steinke2004single, dewan2004review, sheikholeslami2015review, Kaushik2012, Pati2013}]. Apart from enhancement on transport properties, swirling flows in pipes have varied applications such as cyclone separators and swirl atomizers for combustion. Hence, such flows have been very well studied and researched in literature. Swirling flow has often been initiated by introduction of twisted tapes within the tube and sometimes by using swirl generators. Swirling flow in a pipe i.e. flow where an additional velocity in the azimuthal direction is introduced, may decay if there is no continuous forcing in the azimuthal direction. The main reason behind the swirl decay along the length of the pipe is the wall effect, that is the friction between pipe wall and fluid and due to viscous nature of fluid.

Therefore, it becomes very important to study the decay of swirl in a pipe as well as micro-pipe. Perhaps one of the first study on decay of swirl in a tube was experimentally reported by \cite{kreith1965decay}, where the authors developed a correlation for the distance one which the swirl decay as a function of Reynold number for a turbulent pipe flow. Further literature includes work done by analytical, numerical as well as experimental methods. \cite{carrier1971swirling} studied the boundary layer of swirling flow within a rotating container. \cite{kiya1971laminar} solved the swirl decay in a laminar pipe flow numerically using the finite difference method and examined the flow development and entrance zone in detail. \cite{reader1994decay} solved semi-analytically for the swirling flow in a turbulent pipe and found series solution in the form of Bessel functions. \cite{yu1994general} using semi-analytical methods found the swirl decay to be more or less exponential. \cite{Maddahian2011} studied using semi-analytical tools the effect of wall boundary layer on the swirl decay and swirl intensity in a pipe. Correlation for decay of swirl in a laminar pipe flow using numerical techniques was developed by \cite{Ayinde2010}. The author obtained a correlation for the swirl number distribution along the pipe length and found that the swirl number at any location along the pipe length depends on the intensity of swirl number at inlet, the flow Reynolds number, the distance from the pipe inlet, the pipe diameter and the nature of the inlet swirl. \cite{pati2018effects} computationally studied the effect of variation of thermo-physical properties of the fluid with temperature in swirling flow through pipes; the authors quantified the swirl decay as a function of the temperature variation within the fluid. \cite{Kaushik2012a} investigated numerically the decay of swirl in a laminar swirling flow of fluid through a micro-tube with slip at the walls. The authors developed a mathematical correlation for the intensity of swirl at various axial locations along the flow. \cite{Yao2012}, solved analytically the laminar swirl decay in a straight pipe considering no-slip at the walls by using the assumption of fully developed axial flow velocity profile. The authors solved the swirl velocity equation by the separation of variable technique by providing inlet swirl velocity distribution, which includes a forced vortex in the core and a free vortex near the wall (Rankine vortex). The solution which was obtained analytically was compared with the numerical solutions and plotted for different combinations of influential parameters. The authors found that the solution that was obtained analytically depends on the flow Reynolds number, the pipe axial distance, and the inlet swirl intensity. Further, after finding the solution of the swirl velocity distribution, the swirl decay was analyzed.

Due to the recent advances in micro-scale applications such as micro-mixers, micro-separators and micro-combustors, which has created interest among research community, there is a requirement in the research fraternity to enhance transport at the microscale level. At the microscale level the diameter of the tube is in the order of microns hence the Reynold number is typically small, and flow tends to be laminar. Furthermore, in order to increase the mixing in the flow, swirl component of velocity may be introduced. In micro and nano-channels there may be slipping of the fluid at the wall (\cite{neto2003evidence, lauga2003effective, Chakraborty2008, chakraborty2008implications, tretheway2004generating}). The wall slip in surfaces made of carbon nanotubes may be of the order of microns (\cite{majumder2005nanoscale}). Based on these physical considerations, \cite{Kaushik2012a} numerically investigated decay of swirl velocity in a pipe while considering slip at the wall and constant axial velocity profile at the inlet. However, to the best our knowledge there has been no analytical solution for swirl velocity profile and its decay when slip at the wall is considered. Moreover, mixing characteristics of decaying swirling flow with wall slip have not been quantified. Therefore, in our present study we find the analytical solution for swirl decay considering wall slip. The analytical solution obtained is verified numerically. All the assumptions used for analytical solutions are not considered while obtaining numerical solution. A good match is obtained between the numerical solution and the present analytical solution. Further, the decay of swirl and mixing characteristics of the flow are analyzed.

\section{Mathematical Formulation and Solution}\label{sec:math_formul}
\begin{figure}
  \centerline{\includegraphics[width=100mm]{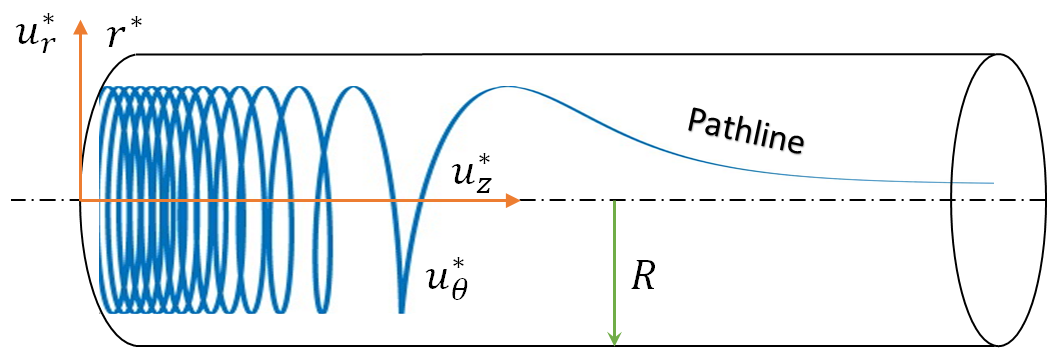}}% Images in 100% size
\setlength{\abovecaptionskip}{0pt plus 0pt minus 0pt}
  \caption{A schematic representation of the flow domain and configuration.}
\label{fig:k1}
\end{figure}

A micro-pipe of radius $R$ is considered as shown in figure \ref{fig:k1}. A parabolic profile is given for axial velocity at the inlet. A swirl velocity is simultaneously introduced as a Rankine vortex (a combination of forced and free vortices) at the inlet. The coordinates are taken from the inlet with $r^*$ being along radius, $z^*$ being along the axis and $\theta^*$ being the azimuthal direction. The fluid is let in to the micro-pipe from the left side with swirl velocity superimposed over the axial velocity and it exits from the right side. The equations governing the fluid flow in figure \ref{fig:k1} is the three-dimensional Navier’s Stokes (NS) equations in cylindrical coordinates for an incompressible fluid.
 
Let $u_r^ *$, $u_\theta^ *$ and $u_z^ *$ be the velocity components in the $r^*$, $\theta^*$ and $z^*$ directions, $\mu$ be the dynamic viscosity of the fluid, $P^*$ be the modified fluid pressure (containing the gravitational body force term), and $\rho$ be the density of the fluid. In order to simplify the governing equations, we assumed that the flow is steady, laminar, and incompressible. Viscosity and density of the fluid are assumed to be invariant. Flow is assumed to be axis-symmetric and the axial direction fluid velocity is assumed to be fully i.e. $u_z^ *  = u_z^ * ({r^*})$ only.

Using the above mentioned assumptions, the continuity equation gives $\dfrac{{\partial ({r^ * }u_r^ * )}}{{\partial {r^ * }}} = 0$. On solving the equation considering a solid impermeable pipe wall, we obtain $u_r^ *=0$ throughout the micro-pipe. The radial momentum equation is simplified to give the value of pressure gradient in the radial direction as a function of azimuthal velocity, given by $\dfrac{{u_\theta ^{ * 2}}}{{{r^ * }}} = \dfrac{1}{\rho }\dfrac{{\partial {P^*}}}{{\partial {r^ * }}}$. If we assume that the magnitude of the swirl velocity as very small, its square becomes much smaller and we may assume the pressure gradient in the radial direction to be negligible. Therefore, there is no further need to solve this equation. In order to solve the azimuthal momentum equation by separation of variables method, we assume ${{{\partial ^2}u_\theta ^ * }}/{{\partial {z^ * }^2}}$ to be very small and therefore neglected. Further, the respective simplification of the azimuthal and axial momentum equations , considering the above given assumptions gives:
\begin{equation} \label{simthetamom_eqn}
u_z^ * \frac{{\partial u_\theta ^ * }}{{\partial {z^ * }}} = \frac{\mu }{\rho }\left[ {\frac{1}{{{r^ * }}}\frac{\partial }{{\partial {r^ * }}}\left( {{r^ * }\frac{{\partial u_\theta ^ * }}{{\partial {r^ * }}}} \right) - \frac{{u_\theta ^ * }}{{{r^{ * 2}}}}} \right]
\end{equation}
\begin{equation} \label{simzmom_eqn}
\frac{1}{\rho }\frac{{\partial p}}{{\partial {z^ * }}} = \frac{\mu }{\rho }\left[ {\frac{1}{{{r^ * }}}\frac{\partial }{{\partial {r^ * }}}\left( {{r^ * }\frac{{\partial u_z^ * }}{{\partial {r^ * }}}} \right)} \right]
\end{equation}

Boundary conditions on $u_\theta ^ *$ i.e. for equation \ref{simthetamom_eqn} are: $u_\theta ^ * (z,0) = 0$, $u_\theta ^*(z^*,R)  = {\rm{ l}}_s^ * {\left| {\dfrac{{\partial u_\theta ^ * }}{{\partial {r^ * }}}} \right|_{r^ *=R}}$ and $u_\theta ^ * (0,{r^*}) = u_{\theta ,i}^ * \left( {{r^*}} \right)$. Here, $l_s^ * $ is the slip length at the pipe wall and $u_{\theta ,i}^ * \left( {{r^*}} \right)$ is the swirl velocity profile at the inlet of the pipe.

In order to solve equation \ref{simthetamom_eqn}, the equation is converted into non-dimensional form by using a reference axial velocity as $U_{ref}$ and length scale as $R$ (radius of the pipe). The dimensionless velocity in the $\theta$-direction is given by $W(z,r) = {{u_\theta ^ * }}/{{{U_{ref}}}}$ and Reynolds number is calculated with respect to the radius as: $Re = {{\rho {U_{ref}}R}}/{\mu }$. Here, the dimensionless coordinates are $r={r^*}/{R}$ and $Z={z^*}/{R}$. Furthermore, we use the dimensionless axial velocity profile as $U(r) = {{u_z^ * }}/{{{U_{ref}}}}$, to get,
\begin{equation} \label{dlessthetamom_eqn}
Re \times U\frac{{\partial W}}{{\partial Z}} = \left[ {\frac{1}{r}\frac{\partial }{{\partial r}}\left( {r\frac{{\partial W}}{{\partial r}}} \right) - \frac{W}{{{r^2}}}} \right]
\end{equation}

The boundary condition of the dimensionless equation \ref{dlessthetamom_eqn} are $W(Z,0) = 0$, ${\rm{W(Z,1)  =  }}{{\rm{l}}_s}{\left| {\dfrac{{\partial W}}{{\partial r}}} \right|_{R = 1}}$ and ${\rm{W(0,r)}} = \dfrac{{u_{\theta ,i}^ * \left( r \right)}}{{{U_{ref}}}}$.

The above equation \ref{dlessthetamom_eqn}, can be solved only if $U(r)$ is known. The solution of equation \ref{simzmom_eqn} is obtained using the boundary conditions $\dfrac{{\partial u_z^ * \left( 0 \right)}}{{\partial {r^ * }}} = 0$ and $u_z^ * \left( R \right) = l_s^*{\left| {\dfrac{{\partial u_z^ * }}{{\partial {r^ * }}}} \right|_{{r^*} = R}}$ to get $U\left( r \right) = \dfrac{{u_z^ * }}{{{U_{ref}}}} = 2a\left[ {b - {r^2}} \right]$, where $a = \left[ {\dfrac{1}{{1 + 4{l_s}}}} \right]$ and $b = \left( {1 + 2{l_s}} \right)$. The dimensionless slip length $l_s$ is given by ${l_s} = l_s^ * /R$. The solution of equation \ref{dlessthetamom_eqn} with the corresponding boundary conditions, using separation of variables is given by
\begin{equation} \label{gen_sol}
W(Z,r) = \dfrac{{{C_n}{e^{\left( { - \dfrac{{\lambda _n^2Z}}{{{\mathop{\rm Re}\nolimits} }}} \right)}}{e^{\left( { - \dfrac{{{\lambda _n}{r^2}\sqrt a }}{{\sqrt 2 }}} \right)}}}}{r}LaguerreL\left( {\frac{{{\lambda _n}b\sqrt a }}{{2\sqrt 2 }}, - 1,\sqrt {2a} {\lambda _n}{r^2}} \right)
\end{equation}

The above solution given by equation \ref{gen_sol} is a series solution and here LaguerreL(a,b, z) is called the generalized Laguerre function and is given by LaguerreL[n,k,r]$=L_n^k\left( r \right)$  [\cite{arfken1999mathematical}]. Further, it is defined as: 
\begin{equation} \label{laguerre_def}
L_n^k\left( r \right) = \frac{{{e^r}{r^{ - k}}}}{{n!}}\frac{{{d^n}}}{{d{r^n}}}\left( {{r^{n + k}}{e^{ - r}}} \right) = \sum\limits_{m = 0}^n {{{\left( { - 1} \right)}^m}\frac{{\left( {n + k} \right)!}}{{\left( {k + m} \right)!\left( {n - m} \right)!}}} {r^m}
\end{equation}

Here $\lambda _n$ is the eigenvalue, given by, ${\lambda _n} = \dfrac{{{a_1}}}{{\left( {1 + 2\dfrac{{{L_{{a_2} - 1}}\left( {{a_3}} \right)}}{{L_{{a_2}}^{ - 1}\left( {{a_3}} \right)}}} \right)}}$ with ${{\rm{a}}_1}{\rm{ = }}\dfrac{{1 - l_s^ * }}{{l_s^ * \sqrt {2a} }}$, ${{\rm{a}}_2}{\rm{ = }}\dfrac{{{\lambda _n}b\sqrt a }}{{2\sqrt 2 }}$ and ${{\rm{a}}_3}{\rm{ = }}\sqrt {2a} {\lambda _n}$.

In order to obtain the coefficient $C_n$, the inlet swirl flow profile is assumed. The swirl velocity profile is assumed to be a Rankine vortex, given in non-dimensional form as $W(0,r) = \left( \begin{array}{l}\dfrac{r}{{{r_t}}},{\rm{    r}} \le {{\rm{r}}_t}\\\dfrac{{{r_t}\left( {1 - r} \right)}}{{r\left( {1 - {r_t}} \right)}},{\rm{ r}} \ge {{\rm{r}}_t}\end{array} \right.$, $r_t={{r_t^ * }}/{R}$ being the dimensionless transition radius.

Using the definition of Rankine velocity profile, we find the value of $C_n$ using the orthogonality condition for the Eigen-functions by the Sturm-Liouville theorem [\cite{kaplan1981advanced, andrews2003mathematical}], as:
\begin{equation} \label{cn_val}
{C_n} = \dfrac{{\int\limits_0^1 {W\left( {0,r} \right){e^{\left( { - \dfrac{{{\lambda _n}{r^2}\sqrt a }}{{\sqrt 2 }}} \right)}}Laguerrel\left( {\dfrac{{{\lambda _n}b\sqrt a }}{{2\sqrt 2 }}, - 1,\sqrt {2a} {\lambda _n}{r^2}} \right)} \dfrac{{ar\left( {b - {r^2}} \right)}}{r}dr}}{{\int\limits_0^1 {{{\left( {{e^{\left( { - \dfrac{{{\lambda _n}{r^2}\sqrt a }}{{\sqrt 2 }}} \right)}}Laguerrel\left( {\dfrac{{{\lambda _n}b\sqrt a }}{{2\sqrt 2 }}, - 1,\sqrt {2a} {\lambda _n}{r^2}} \right)\dfrac{1}{r}} \right)}^2}ar\left( {b - {r^2}} \right)dr} }}
\end{equation}
At this juncture, a dimensionless number called swirl number is defined as the ratio of the axial flux of angular momentum to the axial flux of axial momentum at a particular axial location following \cite{reader1994decay}. Mathematically it is written as:
\begin{equation} \label{swirl_no}
S(Z) = \frac{{\int\limits_0^R {u_z^ * u_\theta ^ * {r^{ * 2}}d{r^ * }} }}{{R\int\limits_0^R {u_z^{ * 2}{r^ * }d{r^ * }} }} = \dfrac{{\int\limits_0^1 {W(Z,r)U(r){r^2}dr} }}{{\int\limits_0^1 {[U(r)]^2{r}dr} }}
\end{equation}
The ratio of swirl intensity at any axial location to the inlet swirl intensity, given by $S(Z)/S(0)$, is used to understand the decay of swirl. Subsequently in section \ref{sec:results_discussion}, the results obtained above is first validated with results existing in literature as well as using numerical methods. Furthermore, results obtained using the present solution are discussed.

\section{Results and Discussion}\label{sec:results_discussion}
\begin{figure}
\vspace*{-12mm}
        \begin{subfigure}[t]{0.5\textwidth}
                \includegraphics[width=1.2\linewidth]{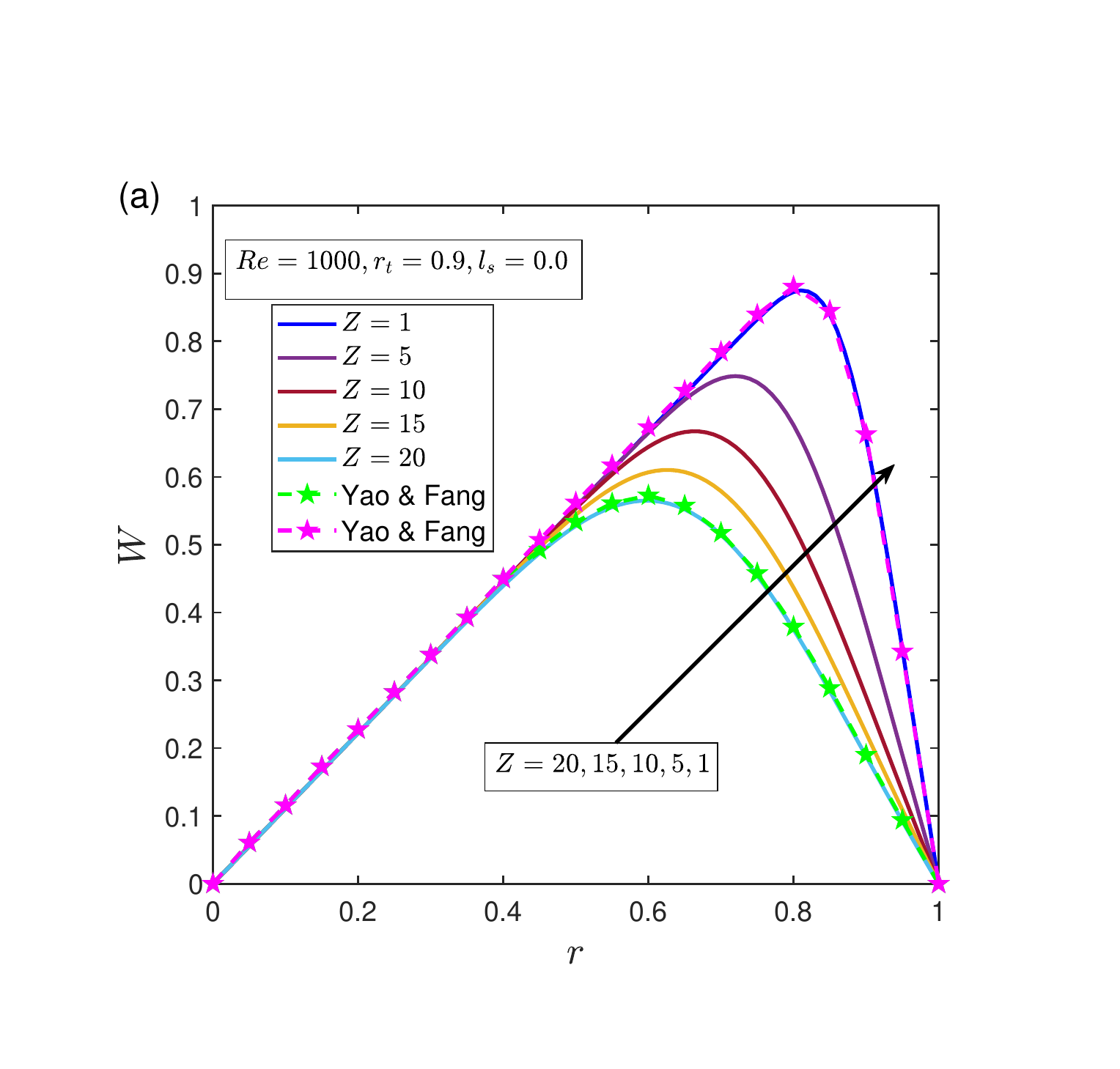}
                \label{fig:k2a}
        \end{subfigure}\hfill
        \begin{subfigure}[t]{0.5\textwidth}
                \includegraphics[width=1.2\linewidth]{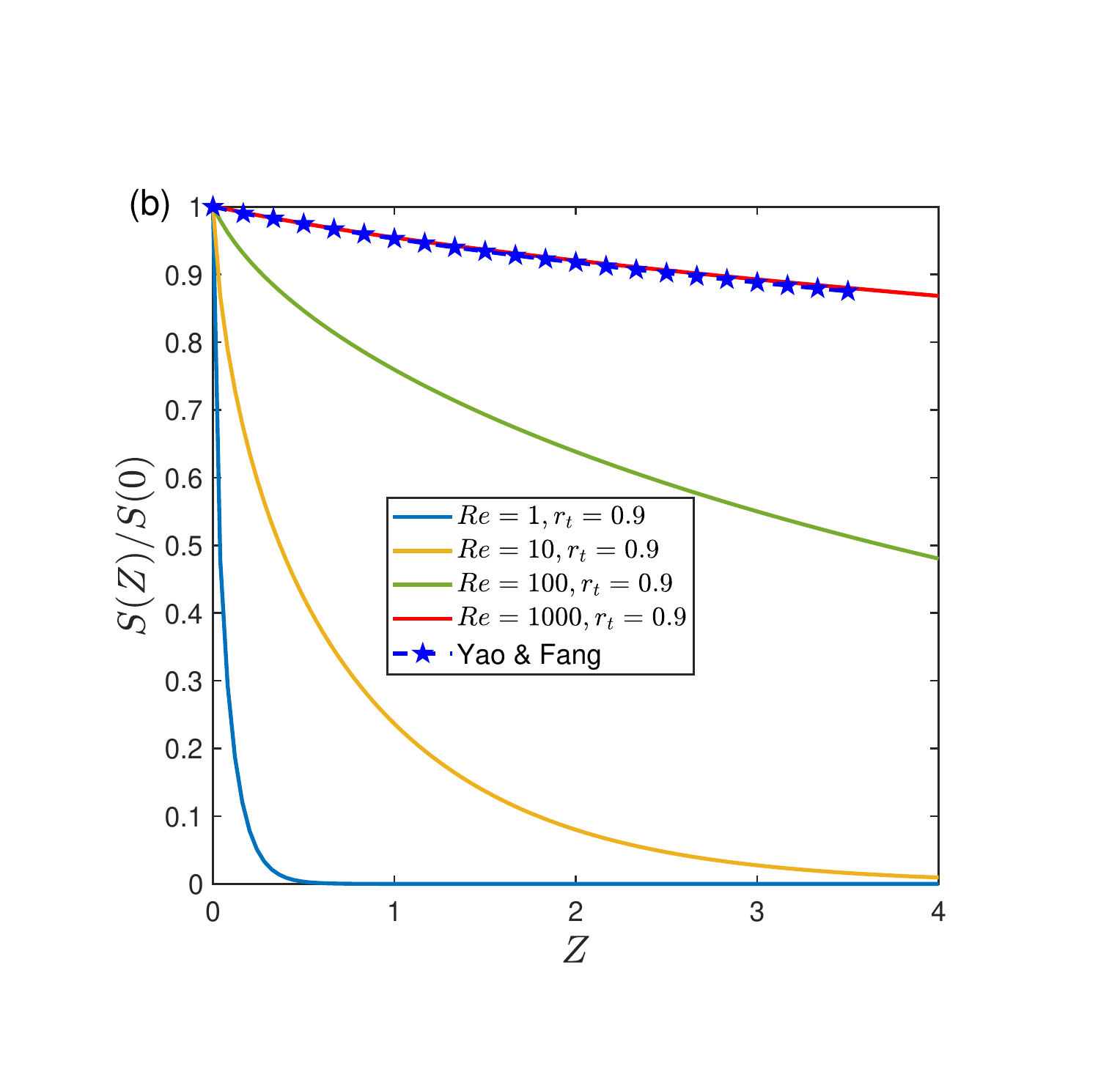}
                \label{fig:k2b}
        \end{subfigure}
\vspace*{-9mm}
 \caption{Validation of (a) swirl velocity profile along radial coordinate and (b) swirl decay in flow without slip along axial distance with results obtained by \cite{Yao2012}.}
\label{fig:k2}
\end{figure}

Before going on to discuss the main results from the present study, it is important to verify the solution found earlier. To check the correctness of the solution obtained in the previous section, two-step validation is performed of the results obtained. First, the results are validated with the results obtained by \cite{Yao2012} for a decaying laminar flow with no slip boundary condition at the wall. For such a case, we set the value of dimensionless slip length $l_s=0$ in the present solution and we obtain the eigenvalues, first five of which are shown in Table \ref{tab:eival}. Also shown in Table \ref{tab:eival} are the first five eigen values for  $l_s=0.1$. 
\begin{table}
  \begin{center}
\def~{\hphantom{0}}
  \begin{tabular}{c|ccccc}
      $n$  & $1$   &   $2$ & $3$ & $4$ & $5$ \\[3pt]
  $\lambda_n$ for $l_s=0$ & 3.2697 & 6.1147 & 8.9484 & 11.7793 & 14.6092 \\
  $\lambda_n$ for $l_s=0.1$  & 3.2083 & 6.0101 & 8.8087 & 11.6124 & 14.4228 \\
  \end{tabular}
  \caption{The first 5 eigenvalues for the generalized Laguerre functionfor 2 different values of $l_s$}
  \label{tab:eival}
  \end{center}
\end{table}
Using the eigenvalues without considering slip at the wall, the solution from the present study is matched with the solution obtained by \cite{Yao2012} and is shown for the swirl velocity profile and swirl decay rate in figures \ref{fig:k2}(a) and (b) respectively. It can be seen from figures \ref{fig:k2}(a) and (b) that the solution obtained from the present study exactly matches with the solution obtained by \cite{Yao2012} when the no-slip boundary condition to the present solution is applied. This is also substantiated by the exact match of the eigenvalues shown in Table \ref{tab:eival} for $l_s=0$ with the Eigen values reported by \cite{Yao2012}. 

\begin{figure}
\vspace*{-12mm}
        \begin{subfigure}[t]{0.5\textwidth}
                \includegraphics[width=1.2\linewidth]{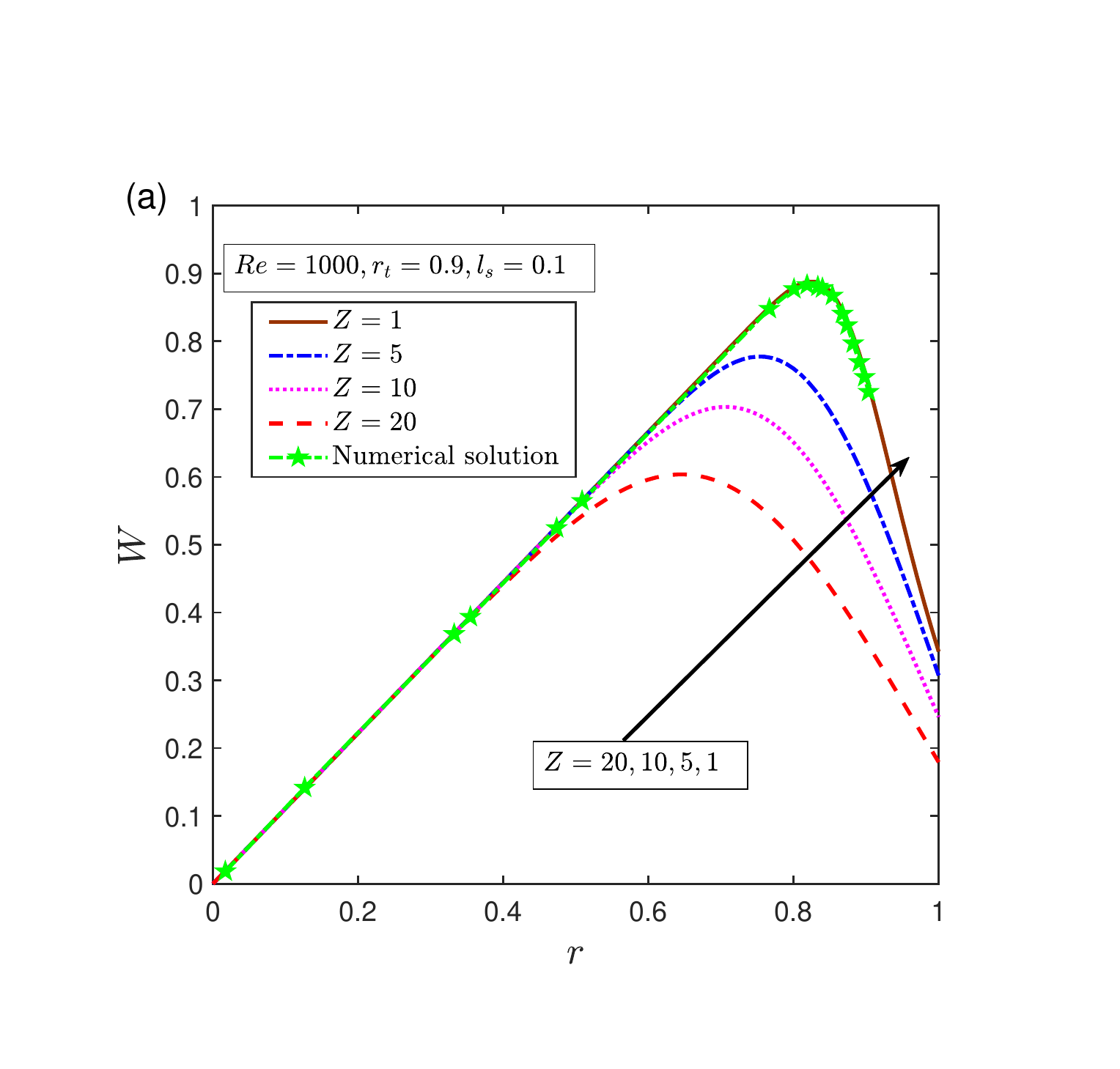}
                \label{fig:k3a}
        \end{subfigure}\hfill
        \begin{subfigure}[t]{0.5\textwidth}
                \includegraphics[width=1.2\linewidth]{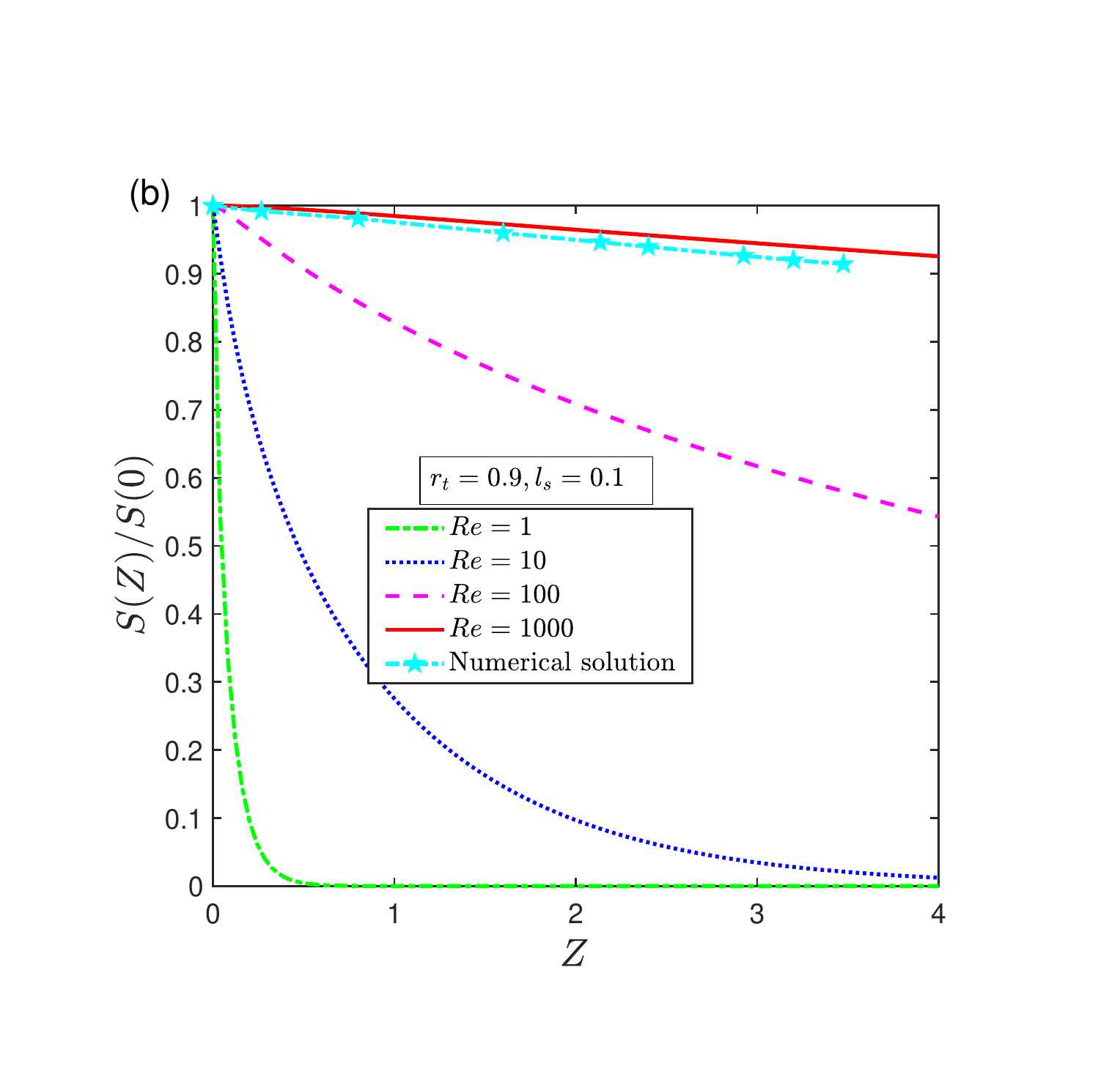}
                \label{fig:k3b}
        \end{subfigure}
\vspace*{-9mm}
 \caption{Validation of analytical solution with numerical solution for (a) swirl velocity profile along radial coordinate and (b) swirl decay along the axial direction, considering with slip $(l_s = 0.1)$ at the wall.}
\label{fig:k3}
\end{figure}

Having validated our solution with the solution obtained by \cite{Yao2012} for the no-slip boundary condition case, it is important to check the correctness of the solution when slip is given at the wall. Therefore, the results obtained from the present study are validated with computational solution obtained using Ansys Fluent 19. This is done in order to find out whether the assumptions made in obtaining the present solution cause it to deviate significantly from the full computational solution. The full momentum equations along the radial, axial and azimuthal directions are solved using Ansys Fluent 19 assuming the axis-symmetricity. The Rankine vortex profile is given for the swirl velocity at the inlet and a parabolic velocity profile (assuming slip at the wall) is given for the axial velocity. Slip is given at the wall as explained earlier. The solution obtained from Ansys Fluent 19 is compared with the present solution for the swirl velocity profile with slip in figure \ref{fig:k3}(a) and with the swirl decay rate in figure \ref{fig:k3}(b). It may be observed from the figures \ref{fig:k3}(a) and (b) that the presentl solution matches very closely with the numerical solution obtained using Ansys Fluent 19. We, therefore infer that our assumptions are sound. 

The discussion on the results obtained from the present study is commenced by examining each parameter that affects the decay of swirl, namely the dimensionless slip length, the Reynolds number and the transition radius. In order to get the results, the eigenvalues need to be calculated for the various slip lengths. As a sample, the first five significant eigenvalues have been calculated and tabulated in Table \ref{tab:eival} for dimensionless slip length $l_s=0.1$. The eigenvalues obtained for $l_s=0.1$ are only slightly different than the eigenvalues obtained for the case with no-slip at the wall, however there is a significant difference in the results of swirl velocity profile and swirl decay for the case with $l_s=0.1$ as compared to the case with no-slip at the wall. This is especially true for the case with larger values of Reynolds number since, in the solution given by equation \ref{dlessthetamom_eqn}, an increase in Reynolds number significantly increases the value of dimensionless swirl velocity for any particular axial location.
\begin{figure}
\vspace*{-12mm}
        \begin{subfigure}[t]{0.5\textwidth}
                \includegraphics[width=1.2\linewidth]{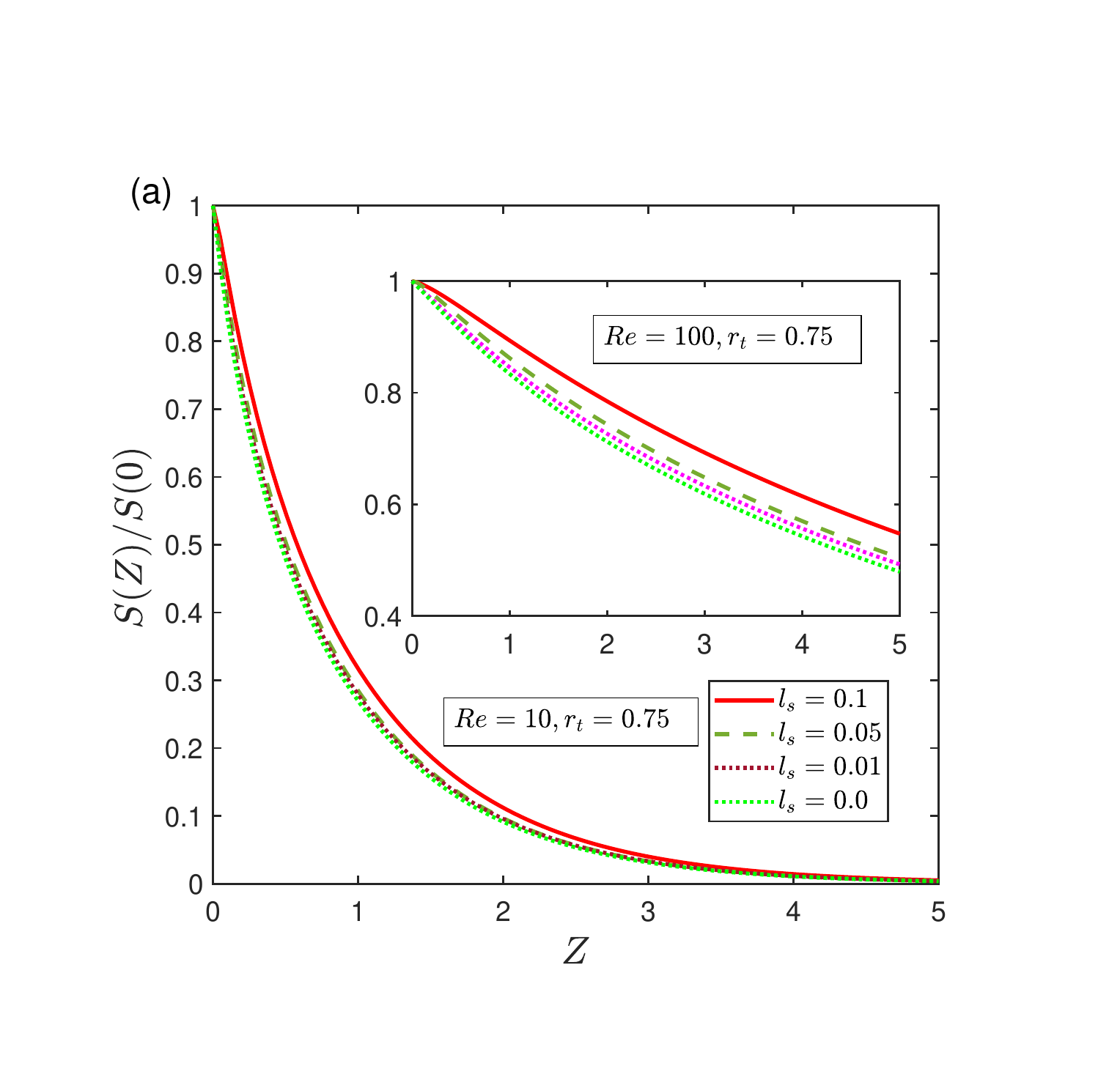}
                \label{fig:k4a}
        \end{subfigure}\hfill
        \begin{subfigure}[t]{0.5\textwidth}
                \includegraphics[width=1.2\linewidth]{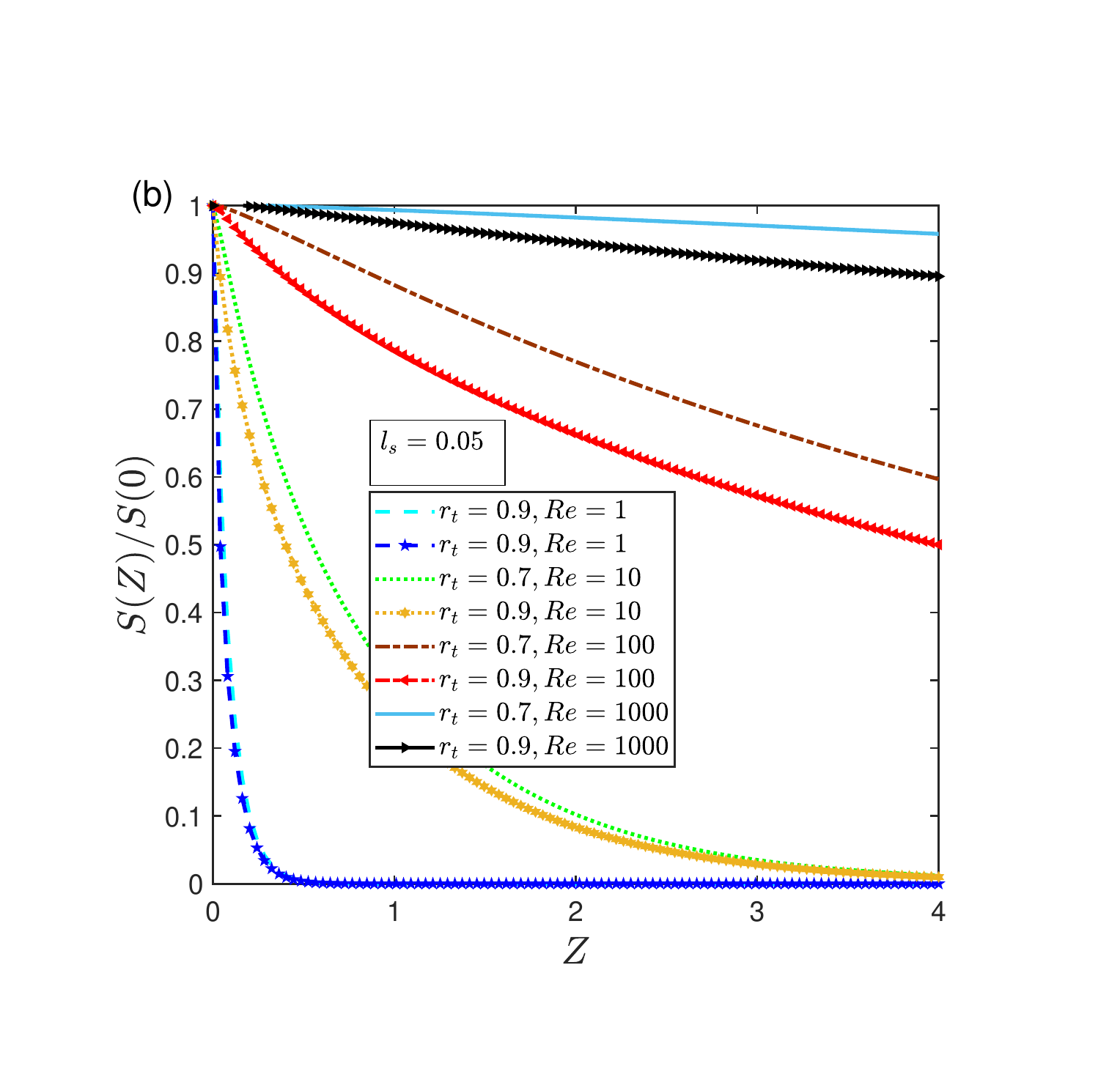}
                \label{fig:k4b}
        \end{subfigure}
\vspace*{-9mm}
 \caption{Swirl decay along the axial location by varying slip length with (a) $Re=50$ and $r_t=0.9$ and (b) at $Re=100$ and $r_t=0.9$.}
\label{fig:k4}
\end{figure}

\begin{figure}
\vspace*{-12mm}
        \begin{subfigure}[t]{0.5\textwidth}
                \includegraphics[width=1.2\linewidth]{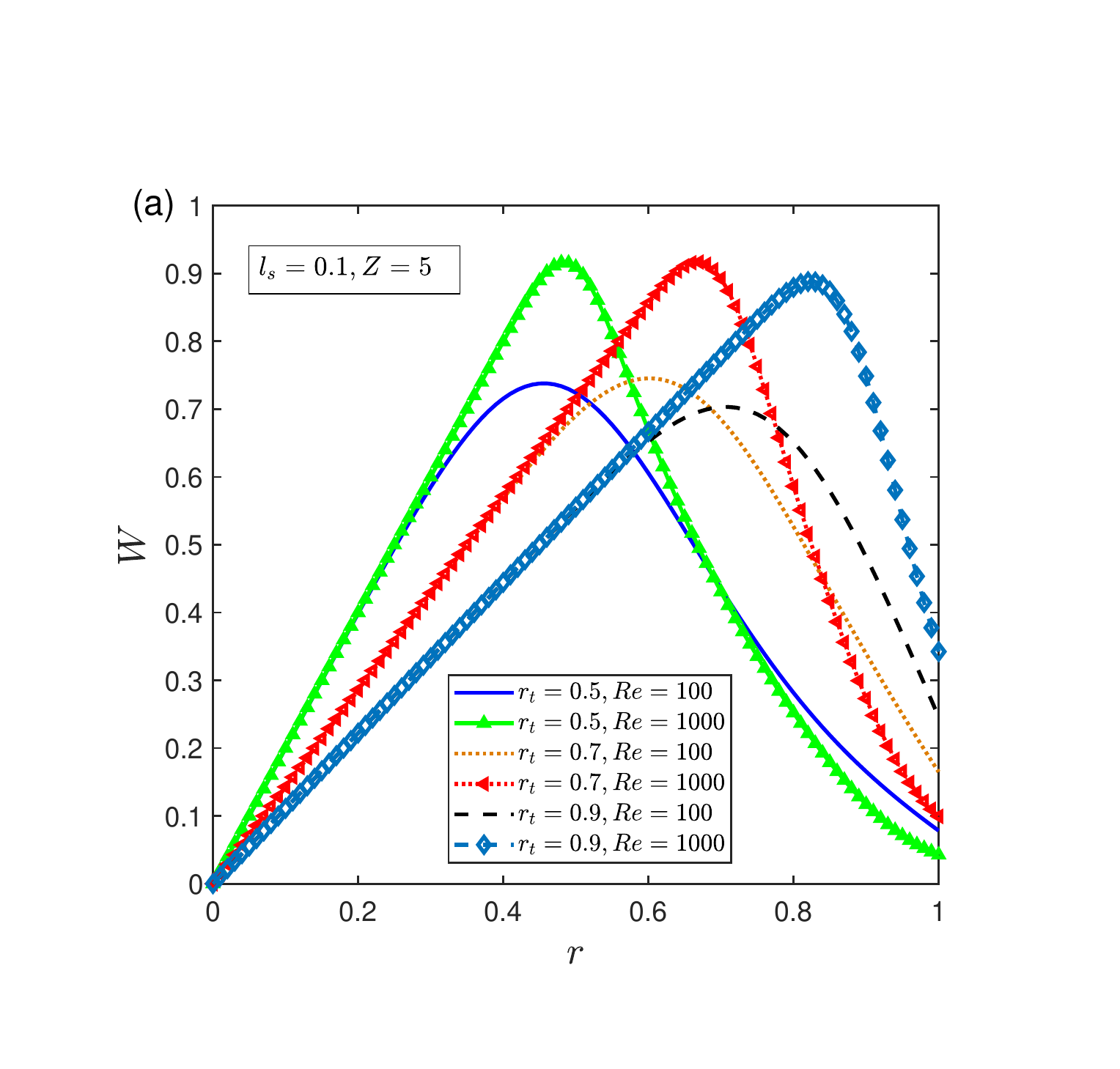}
                \label{fig:k5a}
        \end{subfigure}\hfill
        \begin{subfigure}[t]{0.5\textwidth}
                \includegraphics[width=1.2\linewidth]{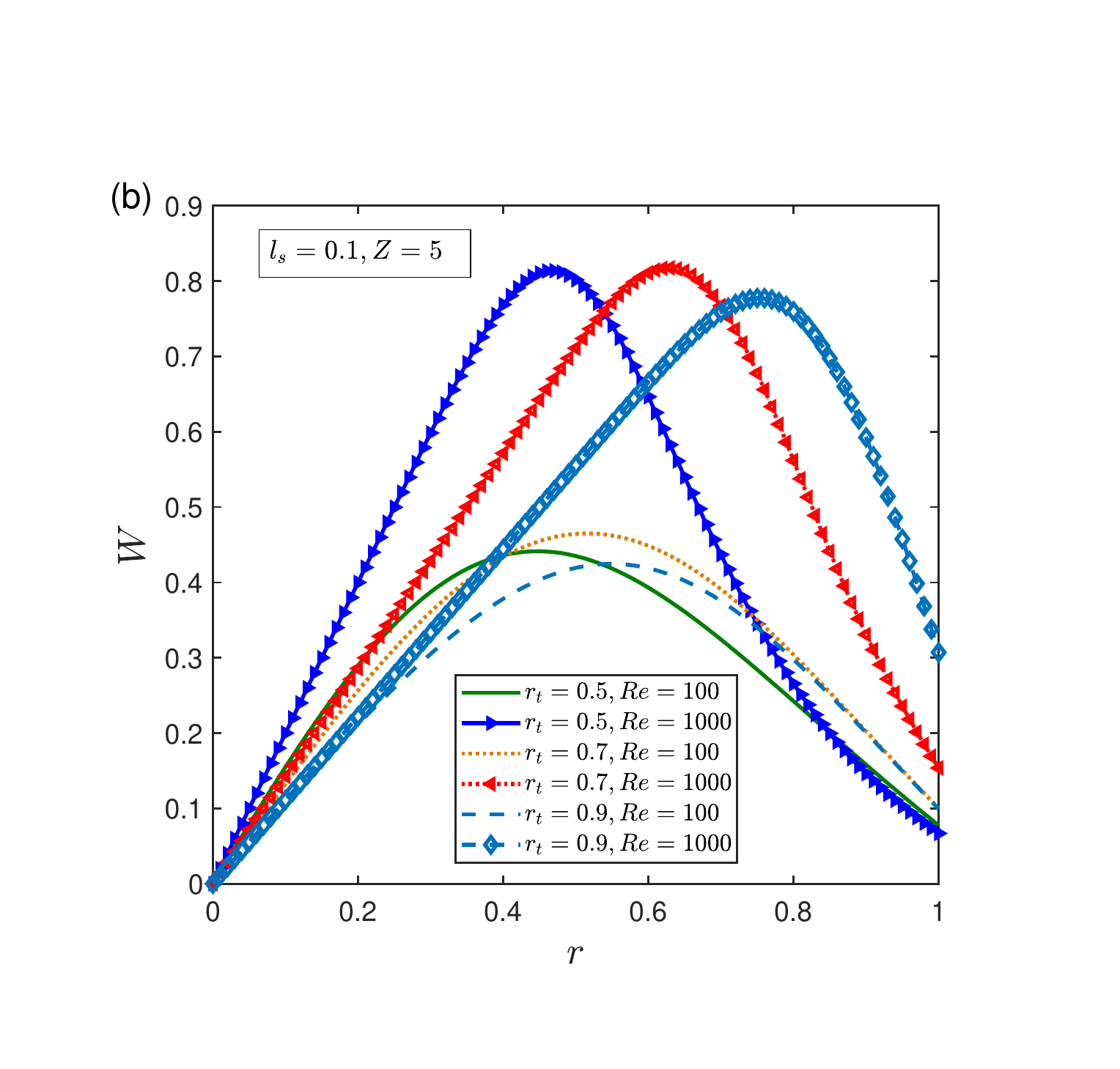}
                \label{fig:k5b}
        \end{subfigure}
\vspace*{-9mm}
 \caption{Swirl decay along the axial location by varying slip length with (a) $Re=10$ and $r_t=0.75$ and (b) at $Re=100$ and $r_t=0.75$.}
\label{fig:k5}
\end{figure}

First, the effect of dimensionless slip length $(l_s)$ is shown in figure \ref{fig:k4}(a) and it's inset. From the figure, it can be seen that the length of swirl decay to obtain similar $S\left( Z \right)/S\left( 0 \right)$ increases with increasing slip length. We see from figure \ref{fig:k4}(a) that irrespective of the value of dimensionless transition radius, the distance to which swirl momentum is transported into the micro-pipe significantly increases with increase in the dimensionless slip length. The primary reason for this is that as the slip length increases the wall shear stress decreases and therefore complete momentum transfer from the stationary walls does not take place to the bulk fluid. It may be also of interest to observe from figure \ref{fig:k4}(a) and it's inset that the for small values of dimensionless slip length, there is very negligible change in the rate of swirl decay. However as $l_s$ increases the change in rate of swirl decay is enhanced. It can be clearly seen from figure \ref{fig:k4}(a) that the effect of $l_s$ in decreasing the decay of swirl is more significant at higher Reynolds numbers.
\begin{figure}
\vspace*{-12mm}
        \begin{subfigure}[t]{0.5\textwidth}
                \includegraphics[width=1.2\linewidth]{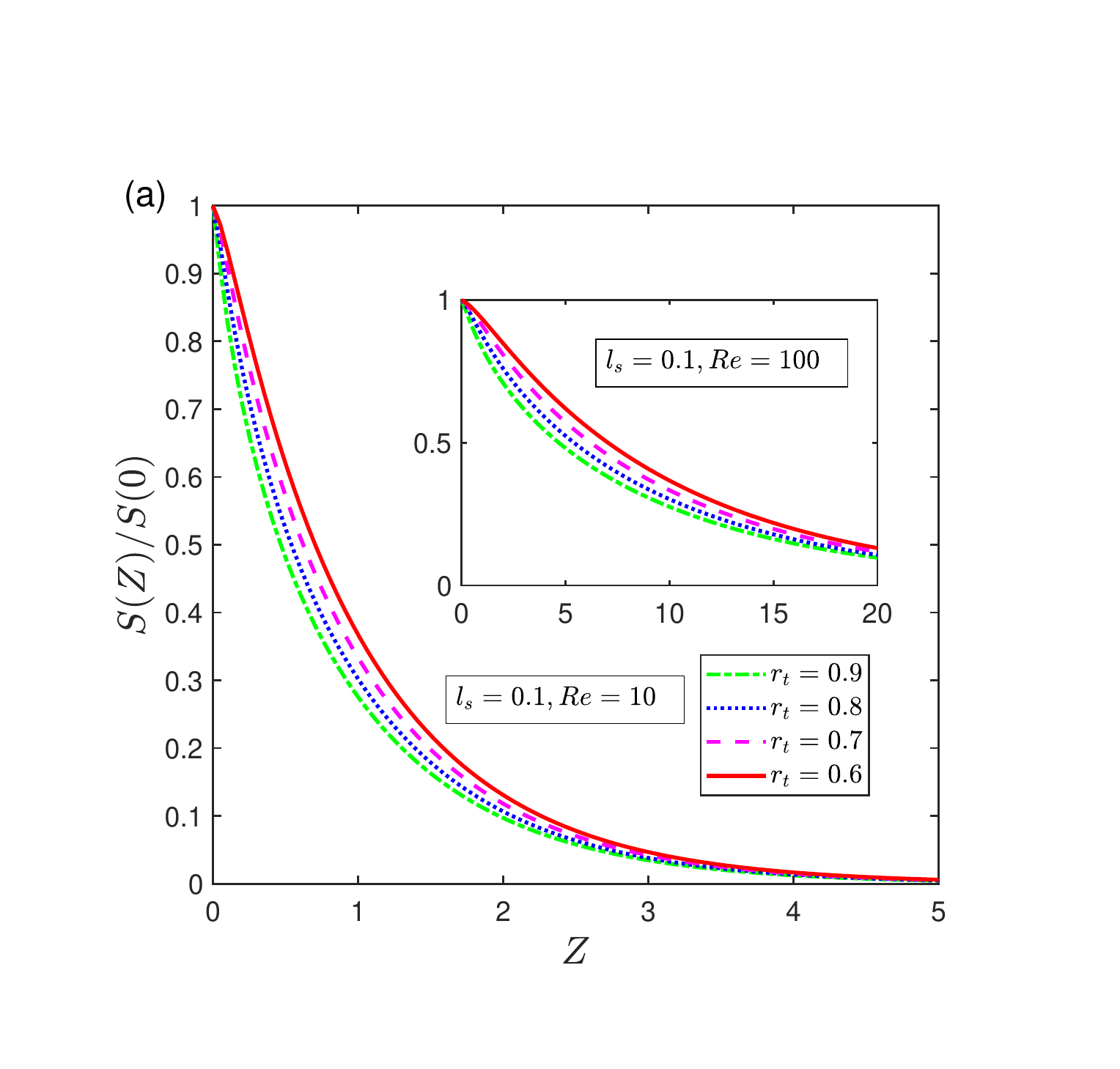}
                \label{fig:k6a}
        \end{subfigure}\hfill
        \begin{subfigure}[t]{0.5\textwidth}
                     \includegraphics[width=9cm,height=9cm]{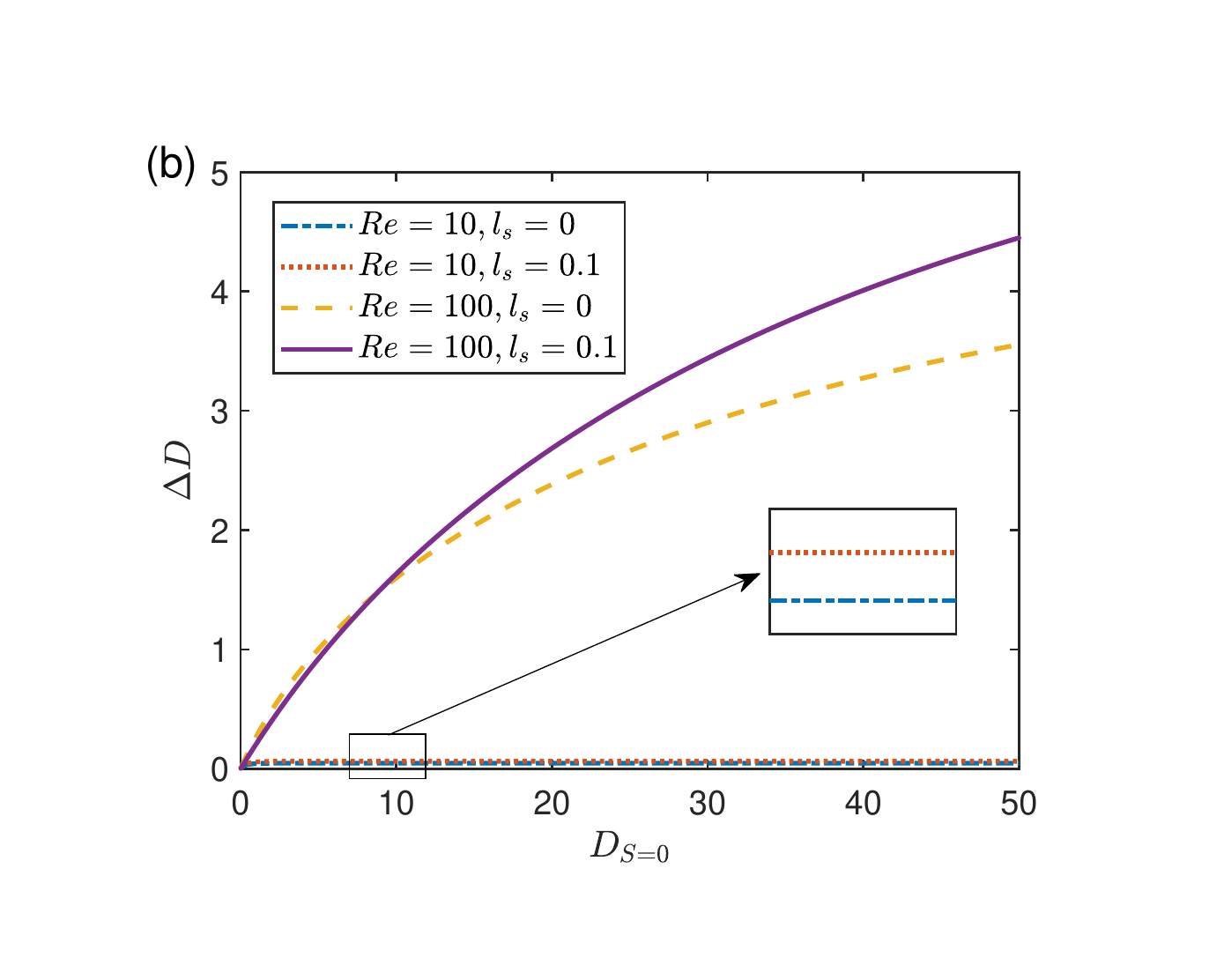}
                \label{fig:k6b}
        \end{subfigure}
\vspace*{-9mm}
 \caption{Decay of swirl intensity by varying the Reynold number having (a) $l_s=0.01$ and $r_t=0.7$ and (b) at $l_s=0.01$ and $r_t=0.9$.}
\label{fig:k6}
\end{figure}

The effect of Reynolds number on the swirl decay is shown in figure \ref{fig:k4}(b). It can be clearly seen that the Reynolds number plays a very significant role in the distance through which the swirl decays. As it can be noted from above solutions, the Reynolds number in the present study is defined using a reference axial velocity, therefore, an increase in $Re$ means the axial velocity is increased. The increased axial velocity causes advection of the swirl momentum farther into the pipe as compared to smaller average axial velocity. This is why it is clearly seen that the swirl decay is significantly lower for larger $Re$ at a particular axial distance from inlet. This trend is true irrespective of the other parameters considered. A good way to decrease the swirl decay may be to increase the Reynolds number along with the dimensionless wall slip.

In order to understand the effect of dimensionless transition radius $(r_t)$ on the swirl decay, we first make an attempt to explain the swirl velocity profile along the radial coordinate when the transition radius is varied. It can be observed from figure \ref{fig:k6}(a) that lower the dimensionless transition radius closer the peak of the dimensionless swirl velocity is to the micro-pipe axis. Since the peak of the dimensionless swirl velocity profile is further away from the wall, the velocity gradients tend to be lower for smaller $r_t$. Due to the smaller value of the gradients, the wall effects have slightly smaller effect on cases with smaller $r_t$. 

The discussion on the effect of $r_t$ on the dimensionless swirl velocity profiles helps us in clearly understanding its effect on the decay of swirl. As we observe from figures \ref{fig:k5}(a) and (b) that the gradients are lower for smaller values of $r_t$, consequently the swirl decay is slower as seen in figures \ref{fig:k5}(a) and (b). The effect of $r_t$ on swirl decay is more pronounced in the case of higher $Re$ as seen in figures \ref{fig:k5}(a) and (b). This is due to the larger axial momentum flux in case of higher $Re$. A good way to decrease the swirl decay may be to increase the Reynolds number and decrease the transition radius along with increasing the dimensionless wall slip.

To understanding the effect of mixing, we analyze the pathline followed by hypothetical fluid particles which are inlet into the microtube at time $t=0$. To find the dimensionless pathline, we integrate the equation $\dfrac{{d\vec x}}{{dt}} = \vec V\left( {r,\theta } \right)$ from $t=0$ to $t$. We then find the dimensionless length of the pathline, which is the dimensionless distance traveled by the fluid particle. We choose many such particles placed equally spaced in the radial direction at the inlet at $t=0$. We keep increasing the number of particles till the average dimensionless distance traveled by these particles becomes almost constant. We denote the average dimensionless distance traveled by these particles by $D_S$, where $S$ represents the swirl number. We further find the difference between the average distance traveled by particles with swirl and without swirl ($D_S=0$) denoted by $\Delta D$. The plot of $\Delta D$ against $D_S=0$ is shown in figure \ref{fig:k6}(b) to show the increase in the average dimensionless distance traveled by the particles when compared to no-swirl flow. The plot also helps in depicting the effect of dimensionless slip length. We observe clearly from figure \ref{fig:k6}(b) that the average distance traveled by the particles in case of swirling flow with wall slip is higher than in the case without wall slip. For both $Re=10$ and $Re=100$ we observe this increase. This clearly indicates the an increase in mixing of swirl-slip interacted flows. 

\vspace*{-5mm}
\section{Conclusion}
In the present study, laminar swirl decay in a straight pipe with wall slip is studied using the analytical solution obtained by solving the governing partial differential equations with slip boundary condition at the wall. The analytical solution is benchmarked with the solution available in the literature for no-slip condition at the wall. The solution is further validated with numerical results, found using finite volume method solver Ansys Fluent. The effects of the Reynolds numbers, wall slip and the inlet swirl profiles on the swirl velocity distribution and decay along the pipe are analyzed and discussed. Results suggest that a high Reynolds number leads to lower decay of the swirl velocity and the swirl intensity and a high transition radius of the inlet swirl profile results in faster decay of the swirl due to the stronger wall drag effects. The results also suggest that the decay of swirl is reduced for larger slip lengths. Higher mixing due to larger average distance traveled by particles is observed for case of swirling flows with wall slip as compared to no-slip condition at the wall. The present work is of significant consequence to micro-pipe studies where enhancement of transport characteristics is sought after.

\bibliographystyle{humannat}  

\bibliography{references}

\end{document}